\documentclass{article}

\usepackage{PRIMEarxiv}

\usepackage[utf8]{inputenc} % allow utf-8 input
\usepackage[T1]{fontenc}    % use 8-bit T1 fonts
\usepackage{hyperref}       % hyperlinks
\usepackage{url}            % simple URL typesetting
\usepackage{booktabs}       % professional-quality tables
\usepackage{amsfonts}       % blackboard math symbols
\usepackage{nicefrac}       % compact symbols for 1/2, etc.
\usepackage{microtype}      % microtypography
\usepackage{lipsum}
\usepackage{fancyhdr}       % header
\usepackage{graphicx}       % graphics
\graphicspath{{media/}}     % organize your images and other figures under media/ folder
\usepackage{amsmath}
\title{Validation and Comparison of HI-STORM Overpack Thermal-Hydraulic Model with MOOSE and NekRS}
\author{
  Sinan Okyay, Elia Merzari, David A. Reger, Victor Coppo Leite\\
        Ken and Mary Alice Lindquist\\ Department of Nuclear Engineering\\
	Pennsylvania State University\\
	University Park, PA 16802\\
  \texttt{sko5200@psu.edu; ebm5351@psu.edu; dzr5281@psu.edu; vbc5085@psu.edu} \\
   \And
  Guillaume Giudicelli, Peter German,Alexander Lindsay\\
        Computational Frameworks\\ 
	Idaho National Laboratory\\
	Idaho Falls, ID 83415 \\
  \texttt{guillaume.giudicelli@inl.gov; Peter.German@inl.gov; alexander.lindsay@inl.gov} \\
}

\begin{document}
\maketitle

\begin{abstract}
Nuclear power is a significant source of electricity in the United States, but the average U.S. nuclear power plant is around 40 years old. Safe management of spent nuclear fuel (SNF) is a crucial aspect of the back end of the nuclear fuel cycle. SNF dry storage systems are increasingly popular as they represent an effective solution in this area, given the absence of a final disposal system. In particular, the spent fuel cask system (dry cask method) provides a feasible solution for maintaining SNF ($\sim$60 years) prior to the final disposal.
\par
The HI-STORM overpack and MPC-32 canister are the primary components of the HI-STORM 100 dry cask storage system. They remove heat from the system via natural circulation with no human intervention required. This characteristic provides passive heat removal while requiring little maintenance in dry cask storage systems.
\par
This project aims to validate and compare the capabilities of a thermal model of HI-STORM overpack developed using the Multiphysics Object-Oriented Simulation Environment (MOOSE) based on the author's previous study. MOOSE is an open-source framework developed by Idaho National Laboratory for multiscale, multiphysics simulations. This study will improve the capabilities of the thermal-hydraulic model of the HI-STORM dry cask storage system by producing high-fidelity results for the air circulation in the overpack. Large Eddy Simulations (LES) are performed using the open-source spectral element code NekRS, developed by Argonne National Laboratory (ANL), for simulating transitional and turbulent flows in complex geometries. NekRS will produce high-fidelity results for the HI-STORM overpack to assess the validity of the current thermal-hydraulic model. 
\par
This work aims to improve the modeling of the HI-STORM overpack system that was previously developed
and validated. The objective of obtaining high-fidelity results is to gain a deeper understanding of the turbulence
characteristics of air circulation in the HI-STORM overpack systems, which will then be used to validate
the assumptions of lower-fidelity models. Finally, the results produced by NekRS will be compared with
those of the current MOOSE model to evaluate the performance of both models.
\end{abstract}

% keywords can be removed

\section{Introduction}

Nuclear power plants have been supplying clean and reliable electricity while offering energy independence for several decades. However, nuclear waste management is one of the challenging topics in the nuclear industry. Spent nuclear fuel (SNF) is the irradiated nuclear fuel that is no longer useful to sustain nuclear chain reactions in a nuclear reactor \cite{Yoo2019}. Even though it is not critical, the spent nuclear fuel (or used nuclear fuel) is highly radioactive and thermally hot due to the radioactive decay of the nuclear fuel. Therefore, it needs to be cooled and safely stored.
\par
Safe management of spent nuclear fuel has been recognized as a critical element of the nuclear fuel cycle \cite{Melcor2020}. Interim storage systems are often preferred in the absence of final disposal or a feasible reprocessing cycle. Interim storage systems offer two ways to store SNF temporarily; wet and dry storage systems. The Fukushima Daiichi accident highlighted the risks of the wet storage method (spent fuel pools) during accidents \cite{Bunn2001}. Therefore, dry cask storage systems are favored to maintain SNF because of the passive cooling mechanisms and low maintenance requirements.
\par
\begin{figure}[ht!]
    \bigskip
    \centering
    \includegraphics[width=1.0\linewidth]{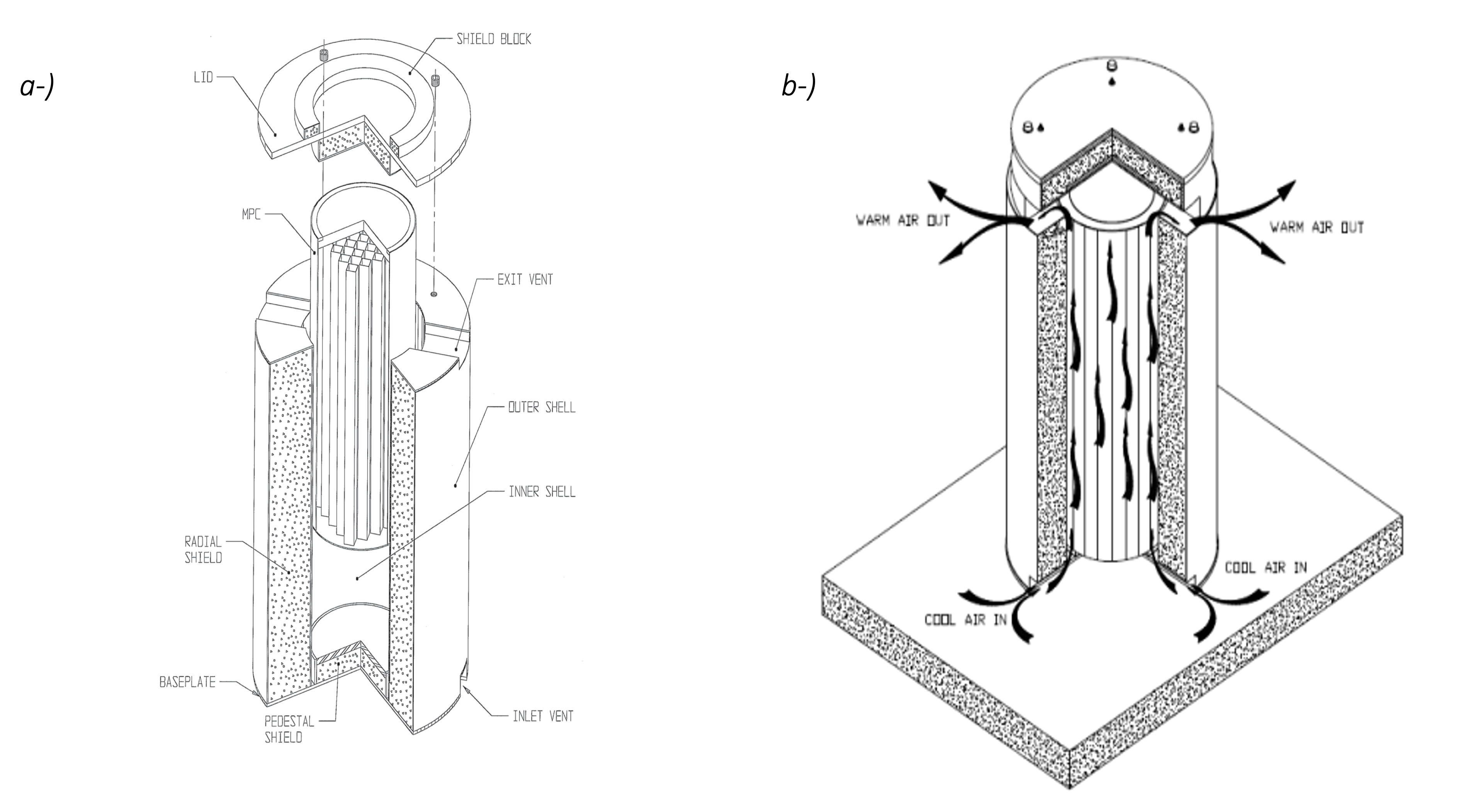}
    \caption{a-) Drawing of the HI-STORM overpack system and b-) Flow patterns of air circulation \cite{Holtec2010}}
     \label{fig:holtec}
\end{figure}

Dry cask storage systems have two main components: the overpack and the multi-purpose canister (MPC). The overpack is a strong-walled cylindrical vessel. The primary shielding function is afforded by plain concrete. The cylindrical concrete vessel is covered with carbon steel for structural purposes, and a thick steel plate is used in the base plate and lid. This study will focus on the overpack part of the HI-STORM 100 dry cask storage system designed by Holtec International. It is pictured in Figure \ref{fig:holtec} from the Final Safety Analysis Report (FSAR) on the HI-STORM system \cite{Holtec2010}.

\par
Thermal-hydraulic characteristics of dry storage systems play a crucial role in the safe cooling of SNF. 3-D Computational Fluid Dynamics (CFD) codes are often preferred to achieve a higher fidelity representation of the systems. \cite{Heng2002,Tseng2011,Herranz2015,Holtec2010}. In order to develop a thermal-hydraulic model of the system, the underlying physical processes should be clearly defined. The heat produced by SNF in the canister is transferred to the overpack air channel through the MPC wall and other structural elements. Eventually, the structural elements, like the wall of the MPC, are cooled down with the natural circulation of air in the overpack. Figure \ref{fig:holtec} shows the natural circulation of air in the system. 

This work aims to improve the modeling of the HI-STORM overpack system that was previously developed and demonstrated using the Multiphysics Object-Oriented Simulation Environment (MOOSE). MOOSE is an open-source framework developed by Idaho National Laboratory (INL) for multiscale, multiphysics simulations \cite{MOOSE2020}. To improve the previous thermal model, higher fidelity levels will be achieved by using NekRS. NekRS is an open-source high-fidelity spectral element code developed by Argonne National Laboratory (ANL), based on the Galerkin discretization similar to finite elements, but with higher accuracy \cite{Fischer2022}.
\par
The results' validation section will focus on the validity of the solvers used in the study. Furthermore, the results of the current thermal-hydraulic model in MOOSE Framework will be demonstrated to show the MOOSE's ability to simulate dry cask scenarios, including natural circulation, heat transfer, and turbulent flow. The objective of obtaining high-fidelity results is to gain a deeper understanding of the turbulence characteristics of air circulation in the HI-STORM overpack systems, which will then be used to validate the assumptions of lower-fidelity models. Finally, the results produced by NekRS will be compared with those of the current MOOSE model to evaluate the performance of both models.

\section{Method}

\par
We used a systematic approach to simulate HI-STORM overpack system. Two numerical models are discussed below in order of complexity. 
The first model focuses on the validation and verification of the results. A differentially heated cavity was chosen as the test case. Cavity systems are often used to measure the validity and performance of computational fluid dynamics (CFD) codes regarding natural convection. Its simplicity and the experimental data available make the cavity system a good candidate for measuring system characteristics. To ensure the validity of the cavity model results, they will be compared with those collected from the other studies \cite{Lo2007}. The validity of NekRS code had been performed in this study \cite{Martinez2020}. Therefore only the validation of MOOSE solver will be presented in the numerical models section for the sake of brevity \cite{Martinez2020}.

The final step focuses on developing a model consistent with the existing HI-STORM overpack. We constructed two separate models from the HI-STORM geometry, a 2-D axis-symmetrical model and a quarter-symmetric 3-D model. The validation of the 2-D axis-symmetrical model will be briefly presented in the numerical models section. This study aims to show differences between the higher fidelity models and the overpack's current thermal-hydraulic model.

\subsection{Major Assumptions and Governing Equations}

\par

The simulation of the HI-STORM overpack is explained in the following sections: the natural circulation of air, material properties, case setup, and   solver. The last section focuses on the settings and summarizes the section.

\subsection{HI-STORM Overpack - Natural Circulation of Air}

Conservation laws dictate the motion of fluids in the continuum limit. The behavior of any fluid can be determined via the mass, momentum, and energy balance equations. The conservation of mass can be expressed as:

\begin{equation}
\frac{\partial \rho}{\partial t}+ \nabla \cdot(\rho \textbf{u}) =0,
\label{eq:Continuity}
\end{equation}

where $\rho$ is the fluid density and $\textbf{u}$ represents the fluid velocity. In natural convection problems, it is often convenient to use the Boussinesq approximation. This approximation assumes that the fluid density changes linearly with sufficiently small temperature differences in the system. Density variations are then neglected in the Navier-Stokes equations but accounted for in the body force gravity term. The Boussinesq approximation can be expressed as:

\begin{equation}
\rho=\rho_0-\beta \rho_0 \Delta T,
\label{eq:Boussinesq Approximation}
\end{equation}

where $\beta$ in the equation represents the thermal expansion coefficient of the system; the time-dependent density term is also neglected in Equation~\eqref{eq:Continuity}. Buoyancy-driven flows can be characterized by the Rayleigh number ($Ra$), a dimensionless number representing the ratio of the time scale due to thermal diffusion and the time scale due to convection. When the Rayleigh number increases, the system becomes more unstable and, eventually, transitions to turbulence \cite{Lo2007}. The Rayleigh number is the product of the Grashof and Prandtl numbers: 

\begin{equation}
Ra=Gr\times Pr= \frac{g\beta \Delta T L^3}{v^2}\times \frac{v}{\alpha} = \frac{g\beta \Delta T L^3}{ v \alpha},
\label{eq:Rayleigh Number}
\end{equation}

where $L$ is the characteristic length of the system. The final form of the conservation equations with the non-dimensional numbers and Boussinesq approximation used in the air-circulation section is as follows \cite{Lo2007}:

\begin{subequations}
\label{eqn:all-lines}
\begin{align}
& \nabla \cdot(\textbf{u}) =0 \label{eqn:line-1}, \\ 
& \frac{\partial \textbf{u}}{\partial t}+ (\textbf{u}\cdot\nabla)\textbf{u} =-\nabla(P)+\frac{Pr}{\sqrt{Ra}}\nabla^2\textbf{u}-Pr T \frac{\textbf{g}}{\textbf{$\mid \textbf {g} \mid$}}\label{eqn:line-2}, \\ 
& \frac{\partial T}{\partial t}+ \textbf{u}\cdot\nabla T = \frac{1}{\sqrt{Ra}} \nabla^2T.\label{eqn:line-3}
\end{align}
\end{subequations}

\subsection{Material Properties}

The air properties in the system are presented in Table \ref{table:property}. Variations of the properties regarding the temperature are small. Therefore, constant properties have been assumed, as this analysis focuses on the steady-state case. These properties' values have been estimated by interpolating the property values found in the FSAR (Final Safety Analysis Report) file of the HI-STORM overpack system\cite{Holtec2010}. 

\begin{table}[ht]
\centering
\caption{Properties}
\resizebox{8cm}{!}{
\setlength{\arrayrulewidth}{0.5mm}
\begin{tabular}{l|c}
 \textbf{Material Properties (400K $\&$ 1atm) $\&$ Constants }     & \textbf{Air}\\
\hline
\multicolumn{1}{c}{\textbf{Name}}  & \multicolumn{1}{|c}{\textbf{Value}}  \\
\hline
Dynamic Viscosity              ($\mu$)  (Pa.s)  & 2.294e-5   \\
Specific Heat                   ($C_p$) (J/kgK)    & 1015     \\
Thermal Conductivity            (k) (W/mK)  & 0.0321 \\
Density                         ($\rho$) ($kg/m^3$)   & 0.8798    \\
Thermal Expansion Coefficient   ($\beta$) (1/K)    & 0.0025 \\
\end{tabular}
\label{table:property}
}
\end{table}
\subsection{Case Setups}
\par
\begin{figure}[ht]
    \centering
    \includegraphics[width=0.6\linewidth]{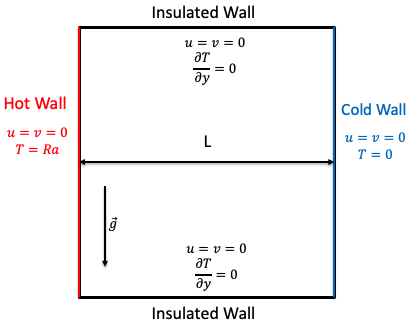}
    \caption{Differentially Heated Cavity case setup.}
     \label{fig:Cavitysetup}
\end{figure}

To construct and mesh the geometries explained previously, MOOSE internal modules \cite{MOOSE2020} and the GMSH code\cite{GMSH2009} were used. The MOOSE modules were used to create the geometry and mesh for the square cavity. A single high-temperature wall in the system triggers the buoyancy-driven flow. Analysis in this step was performed using non-dimensional variables and properties. A Dirichlet boundary condition was applied to the walls, with the left wall being designated as the hot wall with a constant temperature equal to the Rayleigh number and the right wall being designated as the cold wall with a constant temperature of zero. The velocities at the left and right walls were set to zero because they were stationary. The top and bottom walls were considered isolated walls with no-slip boundary conditions. The system's boundary conditions are illustrated in Figure \ref{fig:Cavitysetup}.

The HI-STORM overpack geometry was constructed and meshed using GMSH. GMSH is an open-source 3-D finite element grid generator with a CAD engine to construct the required geometries. It is a fast, user-friendly tool for building desired geometries and meshes \cite{GMSH2009}. The HI-STORM Overpack geometry is modeled with a big air cavity around to capture flow characteristic natural circulation of air. The representation of the geometry can be seen below.

\begin{figure}[ht]
    \centering
    \includegraphics[width=1.0\linewidth]{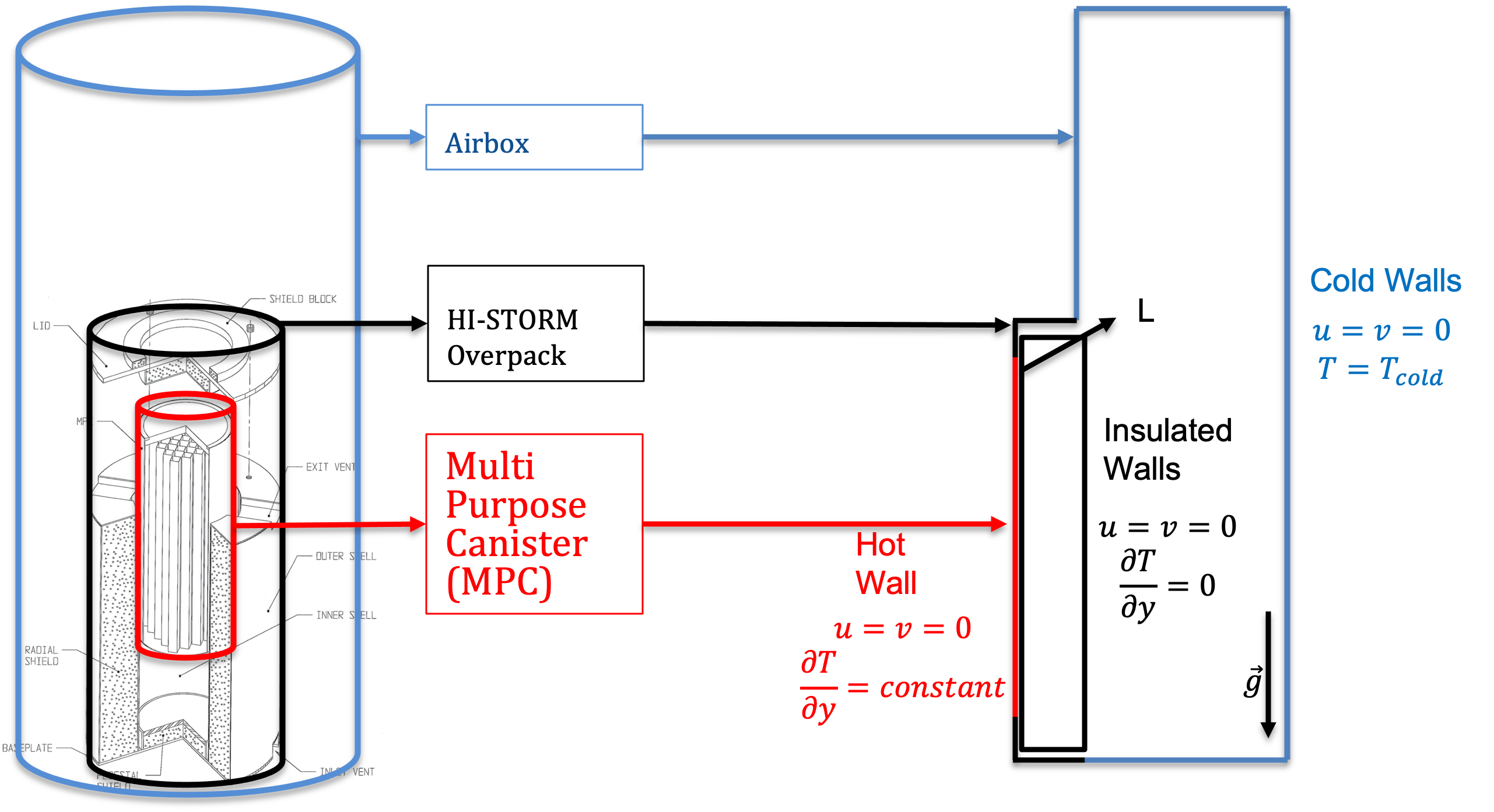}
    \caption{HI-STORM Overpack.}
     \label{fig:HI-STORMOverpacksetup}
\end{figure}

Dirichlet boundary conditions were applied to the cold walls, and constant Neumann boundary conditions were applied through the hot wall to mimic the effects of a multi-purpose canister. The canister was not explicitly modeled in this study, as the focus was on the natural circulation of air. A 2-D axisymmetrical validation of the complete system was performed using the MOOSE Framework solver, and the results will be presented in the results section. 

Additionally, high-fidelity results produced using the NekRS spectral element solver and Large Eddy Simulation (LES) methodology for 3-D quarter symmetry will be generated. The use of LES allows us to assess better the performance of the turbulence models used in the previously developed model. These results will be compared to the current model to evaluate the overall performance of the previously developed model.

\subsection{Solver}
\par
We used MOOSE's Navier-Stokes and Heat Transfer modules to solve the equations described above. The Navier-Stokes module can solve the compressible and incompressible Navier-Stokes (INS) equations via numerical techniques such as Petrov-Galerkin, discontinuous Galerkin, and the finite volume method (FVM). In this study, we demonstrate the solution of the INS equations by using FVM for the natural circulation of air \cite{MOOSE2020}.
\par
Additionally, high-fidelity simulations are conducted on the airbox model to validate the results. Large Eddy Simulations (LES) are performed using the open-source spectral element code NekRS, developed by Argonne National Laboratory (ANL), for simulating transitional and turbulent flows in complex geometries \cite{Fischer2022}. NekRS employs the spectral-element method (SEM), a high-order weighted residual technique that combines the geometric flexibility of finite elements with the rapid convergence and computational efficiency of global spectral methods \cite{Merzari2018}. NekRS offers solutions that require high computational performance and solution accuracy.

The technical aspects of the study are summarized in Table \ref{tab:Table} below.

\begin{table}[ht]
    \centering
    \caption{Summary of the Methods}
\resizebox{12cm}{!} {
\begin{tabular}{ccc}
\textbf{Name}                 & \textbf{Method}         & \textbf{References} \\ \hline
Fluid Flow           & \begin{tabular}[c]{@{}c@{}}Incompressible N-S Equations\\ (Porous \& Non-porous Media)\end{tabular} & \cite{MOOSE2020} \cite{PMOOSE2021}        \\ \hline
Natural Circulation  & Boussinesq Approximation                                                                                      & \cite{Wu2017,Herranz2015}        \\ \hline
Turbulence Resolving & Prandtl's Mixing Length Model                                                                                 & \cite{Versteeg2007}          \\ \hline
Geometry \& Mesh          & Gmsh \& MOOSE Framework                                                                                       & \cite{MOOSE2020} \cite{GMSH2009}          \\ \hline
Solver               & \begin{tabular}[c]{@{}c@{}}MOOSE N-S Module (FVM) \\ NekRS Spectral Element Method\end{tabular}                         &  \cite{PMOOSE2021,Fischer2022}          \\ \hline
% Porous Friction Model   & Darcy-Forcheimer & \cite{Herranz2015} \\ \hline
% Axial Heat Load   & Plateau-Shaped & \cite{Herranz2015,Holtec2010,Wu2017} \\ \hline
% Fuel Thermal Conductivity   & Herranz Derivation & \cite{Herranz2015} \\ \hline
\end{tabular}
}
    \label{tab:Table}
\end{table}

\section{Results}
This section will focus on the cavity model and the outcomes of the HI-STORM Overpack Models.
The following section will show the cavity model results to validate the MOOSE solver's accuracy. The validation of the Nek solver is discussed in the following section. Additionally, a brief summary of the model will be shown to demonstrate the capabilities of the MOOSE solver in complex natural circulation systems. In the second part, a comparison will be made between the airbox and chimney models. The high-fidelity results of the airbox model generated by NekRS will be compared with those generated by MOOSE.

\subsection{Validation}

This part will involve two steps: the validation of the solver and the validation of the HI-STORM model. The solvers will be validated using the differentially heated cavity case. The results of the validation of the MOOSE solver are presented in the next section, and the validation of the NekRS solver is discussed using the study \cite{Martinez2020}. The thermal-hydraulic model for the HI-STORM overpack and MPC-32 were done for a 2-D axisymmetrical model. The results of this have been presented in a previous study by the authors, so the results are briefly summarized to show the validation of the model.

\textbf{\textit{Differentially Heated Cavity}}
\par
\begin{figure}[ht]
    \centering
    \includegraphics[width=1.00\linewidth]{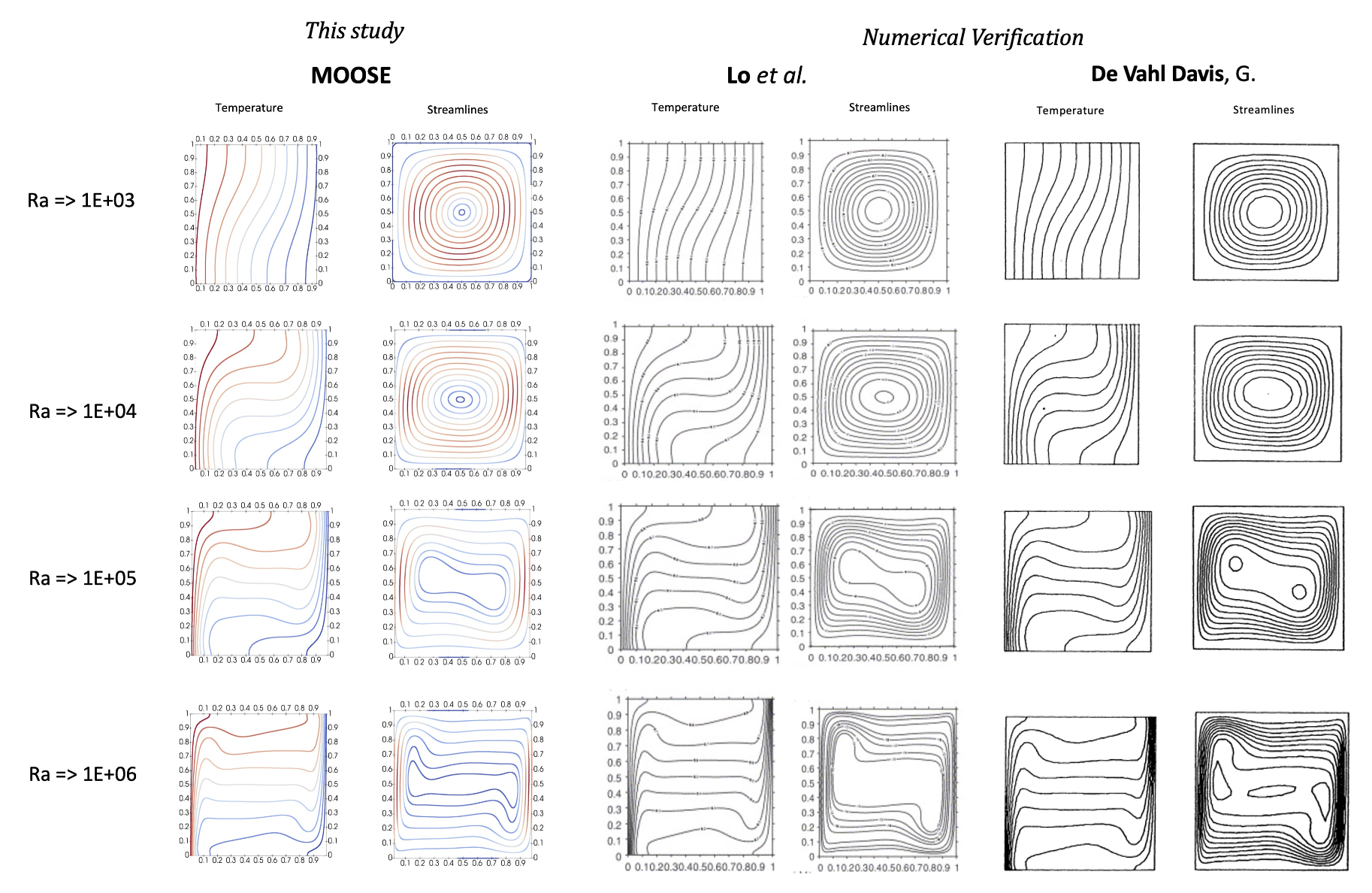}
    \caption{Comparison of isotherms and flow streamlines at several Rayleigh numbers between MOOSE and the work of Lo \cite{Lo2007} and Davis \cite{Davis1983} .}
     \label{fig:cavity}
\end{figure}

The MOOSE solver's accuracy in solving natural convection problems was tested using the well-known cavity problem. The results were evaluated by comparing isotherms and streamlines with published data. The findings are displayed in Figure \ref{fig:cavity}.

The performance of the MOOSE FV N-S solver was evaluated in buoyancy-driven flow problems by comparing the generated isotherms and streamlines with literature for a range of Rayleigh numbers. The results indicate that the MOOSE solver can produce results that are in accordance with the literature. In instances where some small differences were observed in the streamlines for higher Rayleigh numbers, it is believed to be due to a variation in visualization, as the isotherms for the same Rayleigh number were identical with the reference studies.

\par

The cited study \cite{Martinez2020} shows the validation of SEM code using the Differentially Heated Cavity case. The code has been shown to produce results that agree with the literature for the case of a differentially heated cavity. The results of the simulations performed using the Nek code have been demonstrated to have sufficient resolution through a polynomial-order convergence study. The comparison of the first-order statistics with the reference DNS study was found to be satisfactory, with only slight differences observed when examining the fluctuation profiles. The authors attribute these differences to insufficient averaging time in the reference DNS. Overall, the agreement between the two DNS codes validates the numerical method used in the Nek code for the primary situation considered in the study \cite{Martinez2020}.

\textbf{\textit{Validation of HI-STORM Overpack and MPC-32 Model}}
\par
The validation of the HI-STORM overpack and MPC-32 model was performed in the authors' previous study. For the sake of brevity, only the results are presented in this study.
\begin{figure}[ht!]
    \centering
    \includegraphics[width=0.7\linewidth]{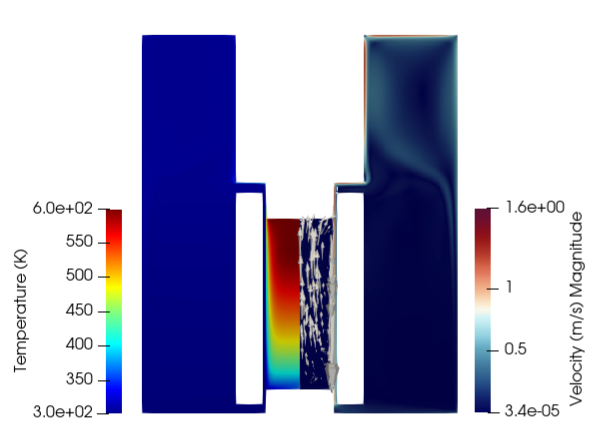}
    \caption{2-D (RZ) simulation results, showing velocity (m/s) at right and temperature (K) at left contours.}
     \label{fig: simulation results}
\end{figure}

Temperature and velocity contours were produced for the given geometry. The air channel's velocity was approximately 1.4 m/s, within the acceptable range based on the reference study. The lower velocity in the air channel may have been caused by the turbulence model used. Additionally, the lower velocity in the airbox model was expected as there were no additional pressure effects from the Bernoulli equation.
 \par

The velocity within the MPC ranges from 1 to 10 cm/s due to the high friction in the porous media. Figure \ref{fig: simulation results} displays the natural flow of helium in the MPC using 2D glyphs. The maximum temperature in the cask, as calculated by MOOSE, is lower than in the reference study, which may be due to the simplified turbulence model, including the approximation of fuel assemblies in the porous medium and ignoring small internal structures in the MPC. These findings are summarized in Table \ref{table: Val 2}. The differences observed are likely a result of the simplification of the axisymmetrical model and the use of a simplified turbulence model.

\begin{table}[ht!]
\caption{Verification of the velocity and temperature predictions of the 2D-RZ model}
\setlength{\arrayrulewidth}{0.25mm}
\begin{center}
\begin{tabular}{ p{3.4cm} |p{1.4cm}|p{1.4cm}|p{1.8cm}  }
\textbf{Name}         & \textbf{Present Study} & \textbf{Herranz et al.\cite{Herranz2015}}  & \textbf{Relative Difference } \\
\hline
MPC Temperature                 & 597 K & 621 K &\textbf{7.47\%} \\ \hline
Air Channel Velocity                 & 1.38 m/s & 1.53 m/s & \textbf{9.80\%}\\
\end{tabular}
\end{center}
\label{table: Val 2}
\end{table}

\subsection{HI-STORM Overpack Large Eddy Simulation (LES) Results }

\par
Our study aimed to investigate the importance of turbulence modeling in the airbox by examining the flow in detail. High-fidelity results were obtained and compared to the N-S module of MOOSE. To ensure mesh convergence, Large Eddy Simulations were performed using the NekRS spectral element solver, with polynomial orders of 5, 6, and 7. Constant heat flux was applied to the air channel of the HI-STORM Overpack.
\par
The numerical model produced the same results for all polynomial orders, indicating sufficient GLL points in the airbox mesh of 1.13 million elements and 246 million GLL points for polynomial order 5. The results were obtained for a 30 kW power system modeled with quarter symmetry. The averaged temperature and velocity distributions of the system are presented below. 

\begin{figure}[ht!]
    \centering
    \includegraphics[width=1.0\linewidth]{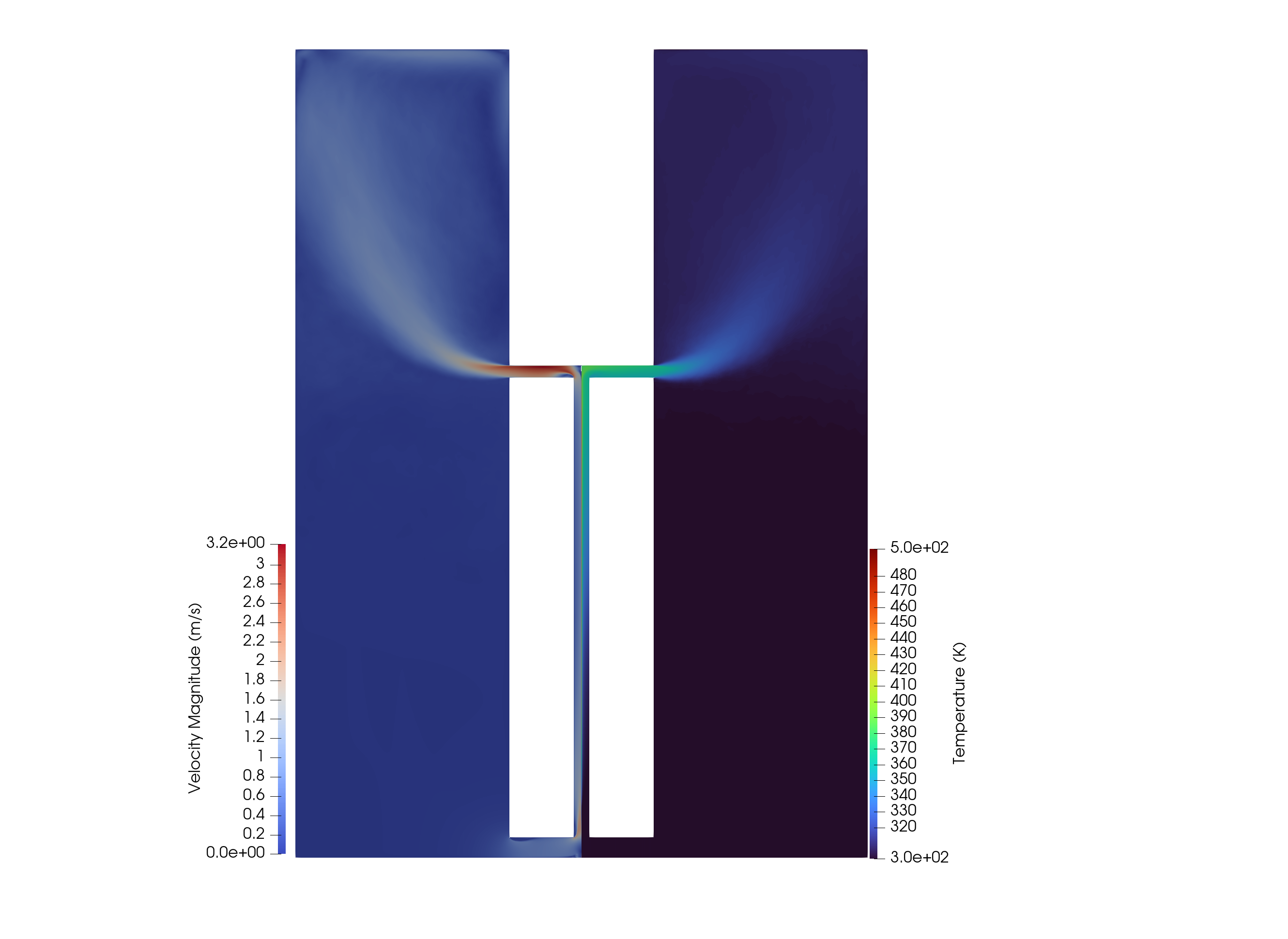}
        \caption{3-D Quarter symmetry LES results, showing velocity (m/s) at left and temperature (K) at right contours.}
     \label{fig: LES Results}
\end{figure}

To make a comparison with the averaged results of coarse CFD Table \ref{table: LES} was produced. The average temperature on the hot wall and average mass flow rate in the air channel were calculated.

\begin{table}[ht!]
\caption{LES Results at Air Channel and MPC Wall}
\setlength{\arrayrulewidth}{0.25mm}
\begin{center}
\begin{tabular}{ p{6cm} |p{2cm}}
\textbf{Name}         & \textbf{Value}  \\
\hline
Average Wall Temperature of MPC  & 463.6 K\\ \hline
Average Mass Flow Rate     &  0.0775 kg/s\\
\end{tabular}
\end{center}
\label{table: LES}
\end{table}

\begin{figure}[ht!]
    \centering
    \includegraphics[width=1.0\linewidth]{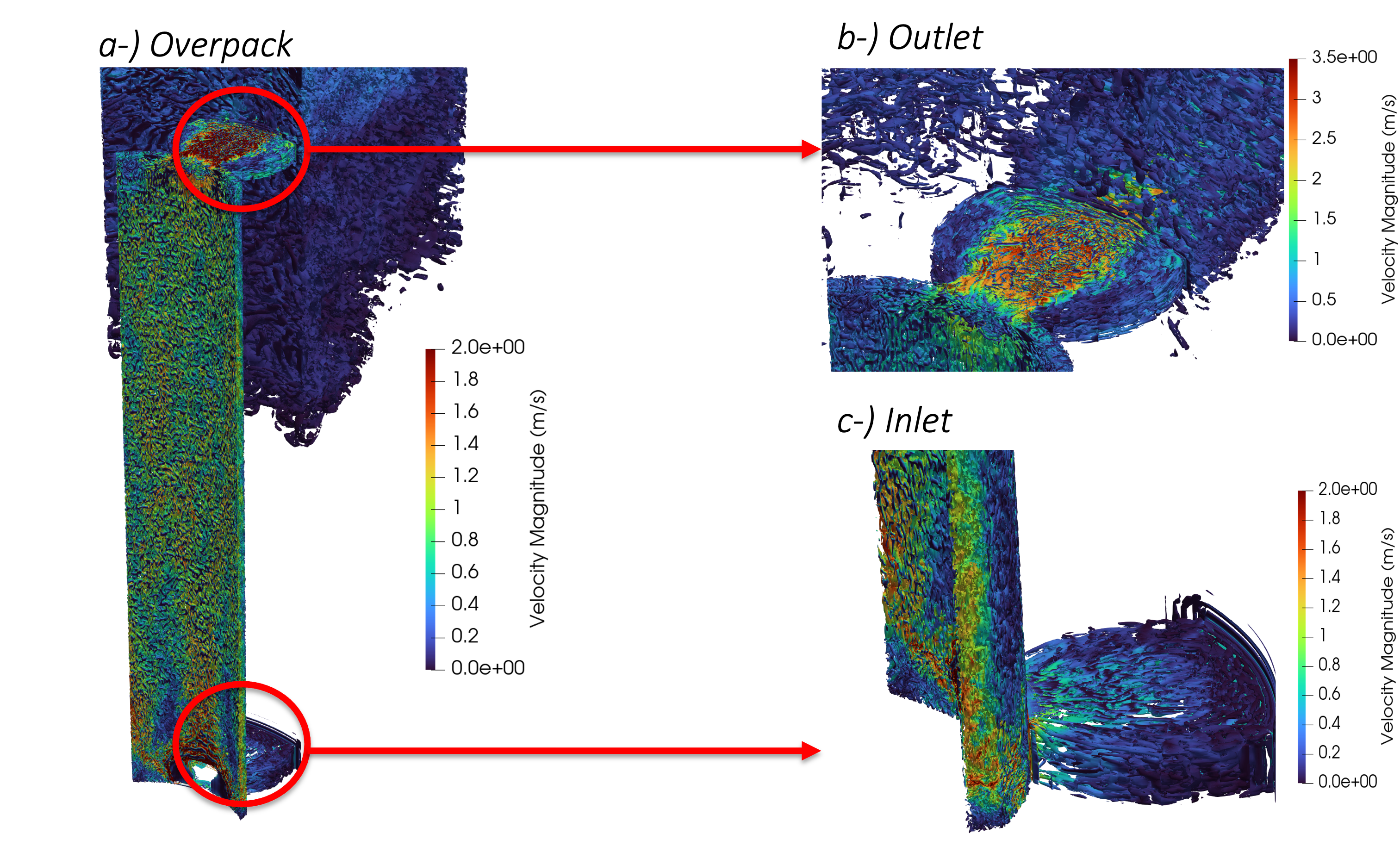}
        \caption{Q-criterion results of the overpack system. The Q-criterion plots are colored with velocity magnitude.}
     \label{fig: qcriterion}
\end{figure}

The Q criterion is used to analyze turbulence in the HI-STORM overpack system. It provides a visualization of the vortical structures in the flow, which are vital characteristics of turbulence. The Q criterion defines vortices as regions where the flow velocity rotates around a central axis. The Q criterion was calculated, and the contours were produced using Paraview. To see detailed structures at the inlet and outlet, separate plots are produced on the right in Figure \ref{fig: qcriterion}. 

The Q criterion plots provide insights into the turbulent field in different regions of the system. In general, elongated structures are identified. This result is expected since the flow configuration consists of a channel, and the structures are generated near the wall. The inlet region is characterized by larger structures that become smaller as the fluid reaches higher velocities in the air channel (higher Reynolds number), where turbulence becomes increasingly dominated by inertial effects. Since this region is where heat is removed, it is generally a  desired outcome. However, while this mechanism ensures higher heat transfer rates, it can increase friction, affecting the overall flow rate. Simulations like this could further improve the overpack's design by shedding light on optimizing the heat transfer rate while minimizing friction losses.

Different from coarser CFD simulations, such as RANS, the LES in this work resolves turbulent structures depicted in Figure \ref{fig: qcriterion}. For this reason, the models developed here do not rely on closures, including Boussinesq turbulent-viscosity hypothesis or the gradient-diffusion hypothesis \cite{Pope2000}. Hence, the results can achieve higher accuracy compared to coarse CFD models.

\subsection{Comparison of MOOSE and NekRS Models}
The results were compared to a 2-D asymmetrical case produced with MOOSE. The instantaneous velocities are compared in the figure below. The MOOSE simulation did not fully represent the buoyant flow due to its overly simplified turbulence model. The results of this comparison highlight the need for better models to capture the turbulence characteristics of the natural circulation in this system. 
\par

\begin{figure}[ht!]
    \centering
    \includegraphics[width=1.0\linewidth]{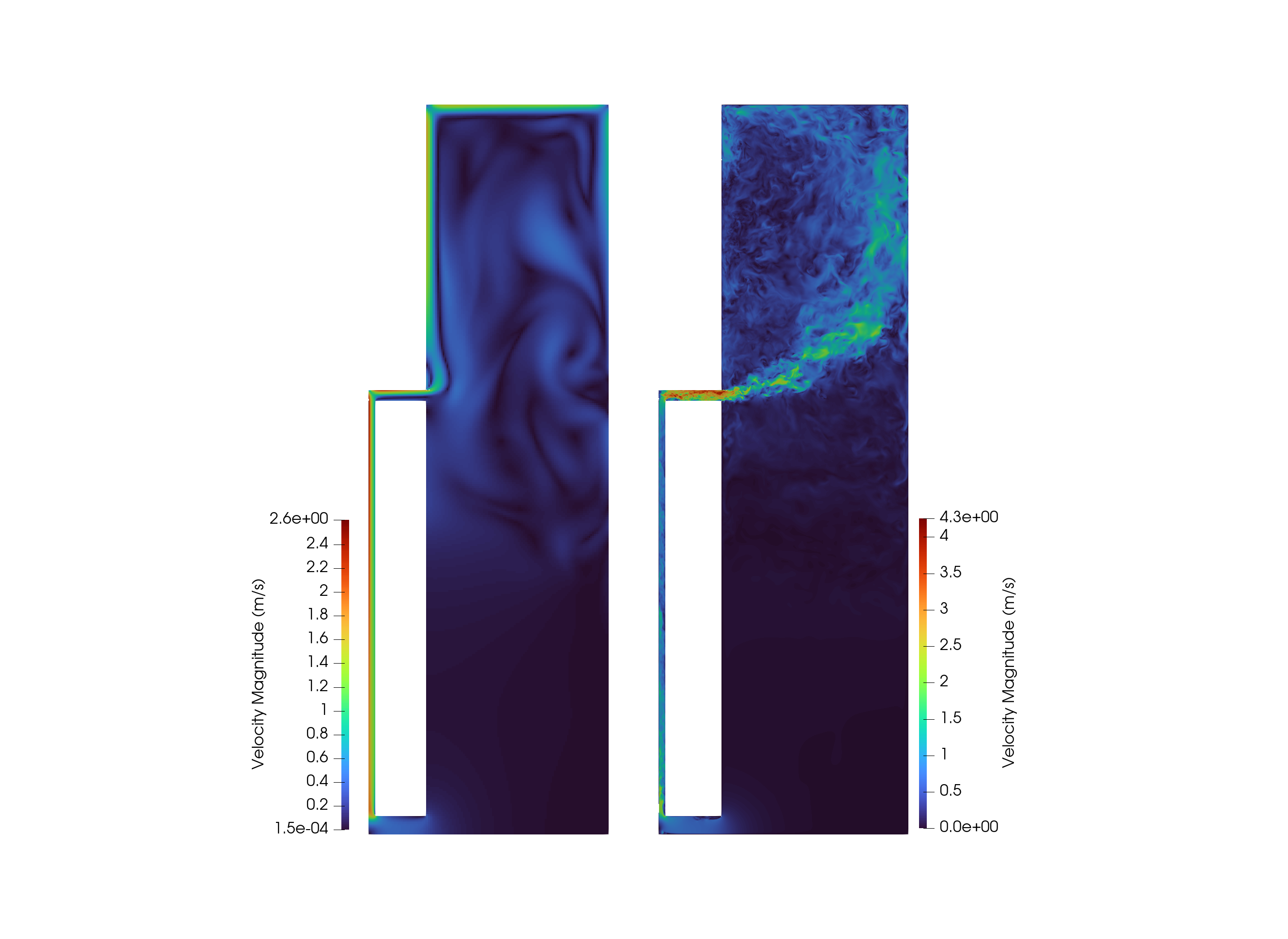}
        \caption{Comparison of velocity distributions between LES (on the right) and coarse CFD (on the left).}
     \label{fig: LESVSCFD}
\end{figure}

\section{Conclusions}

This study aims to evaluate the performance of a previously developed thermal-hydraulic model of the HI-STORM overpack and MPC-32 canister. The validation section of the results presents the validity of the solver used to evaluate the current system's performance. The differentially heated cavity case was used to validate the MOOSE solver, and the results were compared with literature studies. The future study will include turbulence cavity cases for the validation of the MOOSE solver. A previous study validated the SEM solver using a 3-D differentially heated cavity, and the study thoroughly explained the results. The results in this section show that the NekRS and MOOSE solvers can produce the same results with available data in natural circulation problems.

Additionally, the final part of the validation presents the results of the previously developed model of the HI-STORM overpack and MPC-32 canister. These results demonstrate the capabilities of the previous development by comparing it with a reference study. This model can predict the maximum MPC temperature and the peak air channel velocity at certain heights with an error under a certain percentage. The difference can be explained by the axisymmetrical approximation and the simplified turbulence model used in the system. Moreover, the percentage can be caused by the significant reduction in the geometric model of the MPC-32 system.

Finally, the study presents results for high-fidelity simulations, which are of prominent importance in dry cask problems. RANS models are often used to solve dry cask problems, but they may not capture the highly turbulent airflow characteristics in the HI-STORM channel. Q criterion plots provided valuable insights regarding the turbulent field and the heat transfer occurring in the air channel. The same system was modeled using coarse CFD and Large Eddy Simulations to make comparisons. The comparison of the results showed that the coarse CFD simulation did not fully represent the buoyant flow due to its overly simplified turbulence model. This comparison highlights the need for better models to capture the turbulence characteristics of natural circulation in this system. However, it is essential to note that high-fidelity simulations, such as Large Eddy Simulations and higher-order models, may result in more expensive simulations due to increased computational requirements. It is crucial to weigh the benefits of higher accuracy against the computation cost when selecting a simulation method for dry cask problems.

%Bibliography
\bibliographystyle{unsrt}  
\bibliography{templateArxiv}

\begin{thebibliography}{10}

\bibitem{Yoo2019}
Hee~Sang Yoo, Seung~Hun Yoo, and Eung~Soo Kim.
\newblock Heat transfer enhancement in dry cask storage for nuclear spent fuel
  using additive high density inert gas.
\newblock {\em Annals of Nuclear Energy}, 132:108--118, 2019.

\bibitem{Melcor2020}
Michela Angelucci, Luis~E. Herranz, and Sandro Paci.
\newblock Thermal analysis of hi-storm 100s dry cask with the melcor code.
\newblock {\em Journal of Physics: Conference Series}, 1868:012001, 04 2021.

\bibitem{Bunn2001}
M.~BUNN.
\newblock Interim storage of spent nuclear fuel, a safe, flexible and
  cost-effective near-term approach to spent fuel management.
\newblock {\em http://belfercenter.ksg.harvard.edu/files/spentfuel.pdf}, 2001.

\bibitem{Holtec2010}
Holtec Center.
\newblock Holtec international final safety analysis report for the hi-storm
  100 cask system.
\newblock {\em NRC Agencywide Documents Access and Management System (ADAMS)},
  2010.

\bibitem{Heng2002}
Xie Heng, Gao Zuying, and Zhou Zhiwei.
\newblock A numerical investigation of natural convection heat transfer in
  horizontal spent-fuel storage cask.
\newblock {\em Nuclear Engineering and Design}, 213(1):59--65, 2002.

\bibitem{Tseng2011}
Yung-Shin Tseng, Jong-Rong Wang, Fengjee~Peter Tsai, Yi-Hsiang Cheng, and
  Chunkuan Shih.
\newblock Thermal design investigation of a new tube-type dry-storage system
  through cfd simulations.
\newblock {\em Annals of Nuclear Energy}, 38(5):1088--1097, 2011.

\bibitem{Herranz2015}
Luis~E. Herranz, Jaime Penalva, and Francisco Feria.
\newblock Cfd analysis of a cask for spent fuel dry storage: Model fundamentals
  and sensitivity studies.
\newblock {\em Annals of Nuclear Energy}, 76:54--62, 2015.

\bibitem{MOOSE2020}
Alexander~D. Lindsay, Derek~R. Gaston, Cody~J. Permann, Jason~M. Miller, David
  Andr{\v{s}}, Andrew~E. Slaughter, Fande Kong, Joshua Hansel, Robert~W.
  Carlsen, Casey Icenhour, Logan Harbour, Guillaume~L. Giudicelli, Roy~H.
  Stogner, Peter German, Jacob Badger, Sudipta Biswas, Leora Chapuis,
  Christopher Green, Jason Hales, Tianchen Hu, Wen Jiang, Yeon~Sang Jung,
  Christopher Matthews, Yinbin Miao, April Novak, John~W. Peterson, Zachary~M.
  Prince, Andrea Rovinelli, Sebastian Schunert, Daniel Schwen, Benjamin~W.
  Spencer, Swetha Veeraraghavan, Antonio Recuero, Dewen Yushu, Yaqi Wang, Andy
  Wilkins, and Christopher Wong.
\newblock 2.0 - {MOOSE}: Enabling massively parallel multiphysics simulation.
\newblock {\em {SoftwareX}}, 20:101202, 2022.

\bibitem{Fischer2022}
Paul Fischer, Stefan Kerkemeier, Misun Min, Yu-Hsiang Lan, Malachi Phillips,
  Thilina Rathnayake, Elia Merzari, Ananias Tomboulides, Ali Karakus, Noel
  Chalmers, and Tim Warburton.
\newblock Nekrs, a gpu-accelerated spectral element navier–stokes solver.
\newblock {\em Parallel Computing}, 114:102982, 2022.

\bibitem{Lo2007}
D.C. Lo, D.L. Young, and C.C. Tsai.
\newblock High resolution of 2d natural convection in a cavity by the dq
  method.
\newblock {\em Journal of Computational and Applied Mathematics},
  203(1):219--236, 2007.

\bibitem{Martinez2020}
Javier Martínez, Elia Merzari, Michael Acton, and Emilio Baglietto.
\newblock Direct numerical simulation of turbulent flow inside a differentially
  heated composite cavity.
\newblock {\em Nuclear Technology}, 206(2):266--282, 2020.

\bibitem{GMSH2009}
Christophe Geuzaine and Jean-François Remacle.
\newblock Gmsh: A 3-d finite element mesh generator with built-in pre- and
  post-processing facilities.
\newblock {\em International Journal for Numerical Methods in Engineering},
  79(11):1309--1331, 2009.

\bibitem{Merzari2018}
{\em {Performance Analysis of Nek5000 for Single-Assembly Calculations}},
  volume Volume 2: Development and Applications in Computational Fluid
  Dynamics; Industrial and Environmental Applications of Fluid Mechanics; Fluid
  Measurement and Instrumentation; Cavitation and Phase Change of {\em Fluids
  Engineering Division Summer Meeting}, 07 2018.
\newblock V002T09A031.

\bibitem{PMOOSE2021}
Guillaume Giudicelli, Alexander Lindsay, Paolo Balestra, Robert Carlsen, Javier
  Ortensi, Derek Gaston, Mark DeHart, Abdalla Abou-Jaoude, and April~J. Novak.
\newblock Coupled multiphysics transient simulations of the mk1-fhr reactor
  using the finite volume capabilities of the moose framework.
\newblock In {\em Mathematics and Computation for Nuclear Science and
  Engineering}. American Nuclear Society, 2021.

\bibitem{Wu2017}
Yongjia Wu, Jackson Klein, Hanchen Zhou, and Lei Zuo.
\newblock Thermal and fluid analysis of dry cask storage containers over
  multiple years of service.
\newblock {\em Annals of Nuclear Energy}, 112:132--142, 2017.

\bibitem{Versteeg2007}
H.K. Versteeg and W.~Malalasekera.
\newblock {\em An Introduction to Computational Fluid Dynamics: The Finite
  Volume Method}.
\newblock Pearson Education Limited, 2007.

\bibitem{Davis1983}
G.~De~Vahl~Davis.
\newblock Natural convection of air in a square cavity: A bench mark numerical
  solution.
\newblock {\em International Journal for Numerical Methods in Fluids},
  3(3):249--264, 1983.

\bibitem{Pope2000}
Stephen~B. Pope.
\newblock {\em Turbulent Flows}.
\newblock Cambridge University Press, 2000.

\end{thebibliography}

\end{document}